\documentstyle[12pt]{article}

\newcommand{\be}{\begin{equation}}
\newcommand{\ee}{\end{equation}}
\newcommand{\bea}{\begin{eqnarray}}
\newcommand{\eea}{\end{eqnarray}}
\newcommand{\nn}{\nonumber \\}
\newcommand{\p}[1]{(\ref{#1})}
\newcommand{\ba}{\begin{array}}
\newcommand{\ea}{\end{array}}
\newcommand{\vs}[1]{\vspace{#1 mm}}

\def\bbox{{\,\lower0.9pt\vbox{\hrule \hbox{\vrule height 0.2 cm
\hskip 0.2 cm \vrule height 0.2 cm}\hrule}\,}}
\newcommand{\dsl}{\pa \kern-0.5em /}

\newcommand{\pa}{\partial}

\begin{document}

\topmargin 0pt
\oddsidemargin 5mm

\renewcommand{\thefootnote}{\fnsymbol{footnote}}
\begin{titlepage}

\setcounter{page}{0}

\rightline{\small \hfill hep-th/9905029}
\vskip -.75em\rightline{\small \hfill KCL-TH-99-18}
\vskip -.75em\rightline{\small \hfill 5 May 1999}

\vs{15}
\begin{center}
{\Large M-Theory and 
Two-Dimensional Effective Dynamics}
\vs{10}

{\large
N. D. Lambert and P. C. West} \\
\vs{5}
{\em Department of Mathematics\\
King's College\\
University of London\\
The Strand\\
London WC2R 2LS, UK}
\end{center}
\vs{7}
\centerline{{\bf Abstract}}
We calculate from M-theory the two-dimensional low energy effective 
dynamics of various brane configurations. 
In the first part we study configurations 
that have a 
dual description in type IIA string theory as 
two-dimensional $(4,0)$ Yang-Mills theories
with gauge group  $SU(N_1)\times SU(N_2)$ and 
chiral fermions in the bi-fundamental representation. In the second part we
derive related 
equations of motion which describe the low energy internal dynamics
of a supersymmetric black hole in four-dimensional $N=1$ supergravity,
obtained as an M-fivebrane wrapped on a complex four-cycle.

\end{titlepage}
\newpage
\renewcommand{\thefootnote}{\arabic{footnote}}
\setcounter{footnote}{0}

\section{Introduction}

One of the most fascinating aspects of brane dynamics is 
their ability to predict sophisticated results in quantum field theories 
that had previously been studied for many years. 
In particular,  it has been found that one 
can  predict  the  low energy effective actions, including all 
perturbative and non-perturbative effects.  Although these predictions can not 
be made  for a general quantum field theory, 
the cases which can be treated in 
this way are of considerable interest. From the brane perspective 
the non-trivial 
effects often arise from their
classical dynamics and can only be identified with the properties of a
quantum field theory by using some kind of duality.  

In this paper we will use classical brane dynamics to derive 
the low energy effective action for a variety of 
two dimensional quantum field theories. 
In the first part of this paper we focus on theories that have  a 
interpretation within IIA string theory as D-brane configurations and 
hence have a Yang-Mills interpretation. 
In  the second part of this paper we will 
use this technique to derive the low energy effective action for 
four-dimensional black holes which result from M-fivebranes wrapped on a 
four-cycle of a six-dimensional Calabi-Yau manifold. 

Let us consider the situation in \cite{witten} where  brane 
configurations of the form
\be
\matrix{NS5:&1&2&3&4&5& & \cr
         D4:&1&2&3& & &6& \cr}
\label{SWconf}
\ee
were analysed. Here the D-fourbranes end on the NS-fivebranes.
From the standard rules of D-branes the low energy dynamics of this 
configuration is given by a four-dimensional $N=2$ gauge theory.
The technique introduced in \cite{witten} 
was to analyse the strong coupling picture of this configuration
from M-theory. Here the entire configuration appears as a single M-fivebrane
\be
\matrix{M5:&1&2&3&4&5& & & & &\cr
        M5:&1&2&3& & &6& & & &10\cr}
\label{riemann}
\ee
In this paper 
we will consider configurations with less supersymmetry obtained by
adding additional D-fourbranes ending on the NS-fivebranes
\be
\matrix{NS5:&1&2&3&4&5& & \cr
         D4:&1&2&3& & &6& \cr
         D4:&1& & &4&5&6& \cr}
\label{config}
\ee
In M-theory this configuration lifts to
\be
\matrix{M5:&1&2&3&4&5& & & & &\cr
        M5:&1&2&3& & &6& & & &10\cr
        M5:&1& & &4&5&6& & & &10\cr}
\label{hyper}
\ee

The above configurations can be viewed from the worldvolume of the 
first M-fivebrane. There \p{riemann} appears as a threebrane soliton and
\p{hyper} appears as a  string soliton. In this case these configurations
have the interpretation of a single M-fivebrane wrapped around 
a calibrated sub-manifold $\Sigma$ of eleven-dimensional Minkowski 
space~\cite{bbs,glw}.
It was shown in \cite{bbs,glw} that \p{riemann} and \p{hyper}
will preserve one half and one quarter of the M-fivebrane supersymmetry
respectively if the manifold they define is complex. By examining the
form of this manifold appearing in \p{riemann} one can 
derive the Seiberg-Witten curve \cite{sw} for a large class of gauge theories 
\cite{witten}.
In addition, from the equations of motion for 
the M-fivebrane \cite{hsw} one can 
derive a four-dimensional $N=2$ effective theory for the threebrane soliton
\cite{hlw,lw,lwsuper}. This is precisely the Seiberg-Witten low energy 
effective action \cite{sw}, 
including all non-perturbative effects.

Here we will perform a similar analysis of 
the configuration \p{config}. We will show that
it has a two-dimensional chiral $(4,0)$ effective yang-Mills 
theory living on the D-fourbrane intersection. We 
will derive the general form for the M-theory surface \p{hyper}
including all of its moduli.
In addition we will explicitly calculate the effective dynamics  for the 
string soliton \p{hyper} and argue
that it  contains the complete low energy effective dynamics of two-dimensional
Yang-Mills theory.
Note that
M-theory analysis of two-dimensional Yang-Mills theories have already appeared
in \cite{hh,brodie,alishahiha,gu}. However these papers
consider different brane configurations, with $(2,2)$, $(4,4)$, $(4,4)$ 
and $(0,2)$ supersymmetry respectively, and in particular do not calculate 
the low energy effective action. The the effective action for
two-dimensional models with $(2,2)$ supersymmetry were 
obtained  in \cite{en1,en2} using the techniques of geometric engineering
in type II string theory.

One could also view the configuration \p{hyper} in M-theory without
reference to ten-dimensional string theory. 
In this approach, the supersymmetry condition
is interpreted as saying that the M-fivebrane is wrapped on a complex
four-cycle of ${\bf C}^3$. If we consider surfaces that can be compactified
then the string soliton can be reduced to five
dimensions by compactifying M-theory. Wrapping the
string soliton on an $S^1$ and further compactifying to four dimensions we
obtain a black hole in $N=1$ supergravity. Similar black holes were studied
in \cite{msw,vafa} from this point of view and it was found that their
microscopic degrees of freedom correctly accounted for the black hole's
entropy. In this paper we will consider surfaces of this type and derive
the low energy dynamics of the corresponding string soliton. 
This provides a dynamical system of equations for the internal structure of
a black hole at low energies.

The intersecting branes \p{hyper} have also been considered from the point
of view of the AdS/CFT correspondence since the near horizon limit has
an $AdS_3 \times S^2\times {\bf E}^6$ geometry \cite{maldacena}. One
motivation for this work is to help elucidate the relation we between 
description of quantum field theory provided by the M-fivebrane and the
description found by the AdS/CFT conjecture. In general the 
former is not expected to have exact quantitative agreement with quantum
field theory whereas 
the later is. However in this paper we will argue that exact quantitative
information can be obtained from the M-fivebrane. Thus this work should
help explore a configuration where both approaches are valid.

The rest of this paper is organised as follows. In the next section we
analyse the type IIA brane configuration \p{config} using D-branes.
Then in section three we go to the strong string coupling limit of M-theory
and analyse the corresponding configuration \p{hyper}. Next we turn
to the problem of calculating the low energy dynamics of this configuration.
In section four we introduce the M-fivebrane equations of motion and in section
five we derive the two-dimensional effective theory. In section six
we consider a different family of surfaces, whose brane configurations can be 
interpreted as black holes in four-dimensions 
and derive their low energy
effective actions.

\section{The Type IIA Brane Configuration}

Let us first recall the brane configuration \p{SWconf} 
discussed in \cite{witten}.
There one considers two parallel NS-fivebranes in type IIA string theory
lying in the $(x^1,x^2,x^3,x^4,x^5)$ plane and separated along the $x^6$ 
direction by a distance $L$. One then introduces
$N_1$ parallel D-fourbranes in the $(x^1,x^2,x^3,x^6)$ plane,  
suspended between the two NS-fivebranes. The worldvolume theory on $N_1$
parallel D-fourbrane can be described by open string perturbation theory and
at low energy is a five-dimensional $U(N_1)$ gauge theory with sixteen
supersymmetries (the equivalent of $N=4$ in $D=4$). An overall $U(1)$
factor describes the centre of mass motion of the D-fourbranes and is
trivial so we will ignore it. Now consider the presence of the NS-fivebranes.
These are infinitely heavy as compared to the D-fourbranes (which are now
finite in extent along $x^6$). Therefore they do not provide any low energy
degrees of freedom. Instead they cause the low energy D-fourbrane theory to
be reduced to four dimensions and project out half of the supersymmetry. 
Thus the effective theory for the configuration is a four-dimensional
$N=2$ $SU(N_1)$ Yang-Mills theory.

Now we go to the strong coupling limit of the string theory given by
eleven-dimensional M-theory. Here one must introduce an extra dimension
$x^{10}$ which is compactified on a circle with radius $R$. While the 
NS-fivebranes are now simply M-fivebranes the D-fourbranes become M-fivebranes
wrapped around the $x^{10}$ dimension. Thus we arrive at the M-theory 
configuration \p{riemann}. Indeed the whole configuration can be viewed as
a single  M-fivebrane wrapped over a manifold $\Sigma_{SW}$. It was shown
in \cite{witten} that $\Sigma_{SW}$ was precisely the known Seiberg-Witten
elliptic curve appearing in the complete low energy effective action for
$N=2$ $SU(N_1)$ Yang-Mills theory \cite{sw} (see also \cite{klmvw} for
a similar role of the M-fivebrane). In addition by studying the dynamics
of the M-fivebrane one can determine the low energy effective action including
an infinite number of instanton corrections \cite{hlw}. Thus the classical
dynamics of the M-fivebrane is capable of predicting exact coefficients in
four-dimensional supersymmetric Yang-Mills theory.

In this paper we wish to study in a similar way 
the type IIA brane configuration \p{config}
obtained by including $N_2$ additional D-fourbranes to \p{SWconf} in
the $(x^1,x^4,x^5,x^6)$ plane. Thus the intersection between the two
D-fourbranes is a three-dimensional space. 
However the NS-fivebranes insure that
the $x^6$ dimension is finite and hence the low energy theory is 
two-dimensional. In addition the extra D-fourbranes project out another
half of the supersymmetries leaving only four supercharges. 
To see this consider the large string coupling constant limit where  the 
configuration \p{config}
is lifted to the M-theory configuration \p{hyper}. The three M-fivebranes
in \p{hyper} preserve the supersymmetries
\be
\Gamma_{012345} \epsilon = \epsilon \ ,\quad
\Gamma_{0123610} \epsilon = \epsilon \ ,\quad
\Gamma_{0145610} \epsilon = \epsilon \ .
\ee
It follows that the remaining four supersymmetries are chiral
\be
\Gamma_{01}\epsilon = -\epsilon\ ,
\label{susy}
\ee
and hence the two-dimensional low energy theory has $(4,0)$ supersymmetry.

Now consider the the field theory living on the D-brane intersection.
At weak coupling this is obtained by standard D-brane techniques and is
given in \cite{bsv}.
As mentioned above the intersecting D-fourbranes 
reduce the five-dimensional field theory
to three dimensions. Note also that there is a trivial 
$SO(3)\cong SU(2)$ symmetry from the $x^7,x^8,x^9$ dimensions. This leads to
an $SU(2)$ R-symmetry group. 
If we now add the  NS-fivebranes we must dimensionally reduce the
system to two dimensions. The counting of massless states gives the following
modes  in  terms of two-dimensional $N=(4,4)$ 
super-multiplets. From open strings that
begin and end on the D-fourbranes parallel to $(x^1,x^2,x^3,x^6)$ 
we obtain a vector-multiplet with gauge group $SU(N_1)$ 
(here $\mu=0,1$ and $\alpha=1,2$)
\be
a,\chi_A, \quad \lambda^\alpha_{A\pm},\quad A_{\mu}
\ee
where $a = X^4+iX^5$ and $\chi_A= A_2+iA_3$.
There is also a hyper-multiplet 
\be
X^7,X^8,X^9,A_6,\quad \rho^\alpha_{A\pm}
\ee
which is in the adjoint representation of $SU(N_1)$ (since the over-all
centre of mass degree of freedom naturally decouples we have $SU(N_1)$
instead of $U(N_1)$). Here $X^d$ is 
the scalar field representing the fluctuation of the D-fourbrane in the 
$x^d$ direction and $A_m$ is the component of the worldvolume gauge field
in the $x^m$ direction. There are similar
multiplets 
\be
b,\chi_B, \quad \lambda^\alpha_{B\pm},\quad B_{\mu}
\ee
where $b = X^2+iX^3$, $\chi_B = B_4+iB_5$ and
\be
X^7,X^8,X^9,B_6,\quad \rho^\alpha_{B\pm}
\ee
in the adjoint of $SU(N_2)$ found by  examining the open strings
that begin and end on the
D-fourbranes parallel to the
$(x^1,x^4,x^5,x^6)$ plane. So in total we have a vector-multiplet and a 
hyper-multiplet in the adjoint of $SU(N_1)\times SU(N_2)$. Note that these
states are simply the dimensional reduction of ten-dimensional $N=1$
super-Yang-Mills-multiplets reduced to two dimensions. These two multiplets
are by themselves insensitive to each other and
hence preserve the full $(8,8)$ supersymmetry (i.e. the same as a single
D-fourbrane).

In addition there is
a hyper-multiplet
\be
\phi_1,\phi_2,\phi_3,\phi_4,\quad \psi_\pm^\alpha
\ee 
coming from open strings stretching between the two types
of D-fourbrane \cite{bsv}. These states transform in 
the fundamental representation of $U(N_1)\times
U(N_2)$. Again, since a $U(1)\times U(1)$ factor of this group is trivial, 
the interacting part of the hyper-multiplet 
is in the fundamental representation of
$SU(N_1)\times SU(N_2)$. These states  mediate the interaction
between the two factors of the gauge group (i.e. the interactions between the 
two types of D-fourbranes) and break the supersymmetry
to $(4,4)$.

The NS-fivebranes do not just dimensionally reduce the three-dimensional
intersection to two dimensions.  
It was pointed out in \cite{hw} that they also impose boundary
conditions where the D-fourbranes meet the NS-fivebranes which can eliminate
zero modes and break additional supersymmetries. 
Following this argument let us  
consider the adjoint hyper-multiplets. These fields represent fluctuations
of the D-fourbranes in the directions transverse to the NS-fivebranes and,
since the D-fourbranes must end on the NS-fivebrane, we must impose Dirichlet
boundary conditions so that the scalar fields $X^7,X^8,X^9$ vanish at 
$x^6 = 0,L$. This implies that they are
massive modes (with massive of order the string scale) 
and hence they do not appear in the low energy Yang-Mills 
theory. This argument applies separately to each type of D-fourbrane,
as in the construction of \cite{witten}, and so it  follows that the 
$(4,4)$ superpartners $A_6,B_6,\rho_A,\rho_B$ are massive too. 
On the other hand the vector multiplets  represent the
fluctuations of the D-fourbranes within the NS-fivebranes and need not
vanish at $x^6 = 0,L$. Instead we simply impose Neumann boundary conditions
which leave zero modes for all the fields in the $(4,4)$ multiplet.

The boundary conditions imposed on the bi-fundamental
hyper multiplet are more
difficult to analysis since they do not have a straightforward geometrical
interpretation. To this end a more formal argument can be constructed
which reproduces the above massless modes for the adjoint multiplets and also
shows that only the two lowest helicity zero modes of the bi-fundamental 
hyper multiplet remain and are singlets under supersymmetry \cite{lwnew}, 
i.e. two helicity $-{1\over2}$ fermions. 
To summarise,  the surviving two-dimensional zero modes are 
a $(4,4)$ vector multiplet 
$(a,\chi_A,\lambda^\alpha_{A\pm},A_{\mu})$
with gauge group $SU(N_1)$, a second $(4,4)$ vector multiplet
$(b,\chi_B,\lambda^\alpha_{B\pm},B_{\mu})$ with gauge group $SU(N_2)$ 
and two
right-handed fermions $\psi^a_-$ 
in the fundamental of $SU(N_1)\times SU(N_2)$. These
right-handed fermions are singlets under spacetime supersymmetry leaving  
$(4,0)$ supersymmetry on the two-dimensional intersection.

Note that the vector multiplet fields can carry momentum off the intersection
(i.e. in the $x^2,x^3,x^4,x^5$ directions), although the bi-fundamental 
fermions only propagate along $x^0,x^1$. Without the
bi-fundamental fermions the two gauge groups do not interact and 
can equally be viewed as a single four-dimensional $N=2$ 
$SU(N_1)\times SU(N_2)$ gauge theory.
Therefore  the resulting
effective dynamics we obtain more accurately 
describe a one-dimensional 
defect in a four-dimensional $N=2$
$SU(N_1)\times SU(N_2)$ gauge theory. However, since the interactions 
between the two different gauge groups propagate only in two dimensions,
the effective dynamics we obtain should contain the full description of
an interacting $(4,0)$ $SU(N_1)\times SU(N_2)$ two-dimensional gauge theory.

Let us consider first the theory on the first D-fourbrane in the 
$(x^1,x^2,x^3,x^6)$ plane. The NS-fivebranes cause the $x^6$ dimension
to be compact with length $L$. Therefore
the four-dimensional 
Yang-Mills coupling constant $g_{YM}$ is given by the 
ratio $g^2_{YM} = g_s/L$ where $g_s = R^{3/2}$ 
is the string coupling constant \cite{witten}.
It follows that we may go to the large distance, small
curvature limit $R,L\rightarrow\infty$ keeping the couplings fixed. In this
limit M-theory is well described by eleven-dimensional supergravity.
Thus the effective description that we obtain from M-theory should be
accurate for any value of the Yang-Mills coupling constant. In fact we will
see that, as is the case for the $N=2$ theories, the parameter $R$  
does not appear in the low energy dynamics, which are therefore insensitive
to the extra dimension. Another important
restriction on the low energy terms is that they are holomorphic. One may
therefore expect some kind of non-renormalisation theorem that ensures they
are the correct terms for the Yang-Mills theory.
Thus we can
conjecture on the basis of type IIA/M-theory duality that we will arrive
at the exact quantum low energy effective action.

\section{The M-fivebrane Geometry}

Let us now pass to the strong coupling M-theory picture. The D-fourbranes
then lift to M-fivebrane configuration \p{hyper}. Indeed, 
as in \cite{witten},
all branes are M-fivebranes which can in turn be described by a single
self-intersecting M-fivebrane with a complicated worldvolume. The intersection
\p{hyper} can then be viewed as a supersymmetric string 
soliton on the worldvolume  of the first M-fivebrane. Our approach is
to derive the dynamics of the D-fourbrane intersection gauge theory by 
calculating the classical low
energy dynamics of this string soliton of the M-fivebrane.

To perform the M-theory analysis it 
is useful to introduce the complex coordinates
\be
s = X^6 + iX^{10}\ , \quad t = e^{-s/R}\ , 
\quad w = x^2+ix^3\ , \quad z = x^4+ix^5\ .
\ee
It was shown in \cite{glw} that the soliton \p{hyper} is a solution to 
the M-fivebrane equations of motion and preserves one quarter of the
sixteen supersymmetries if $s(z,w)$ is a holomorphic function. 
This has the
interpretation of wrapping the M-fivebrane around a four-dimensional
complex submanifold $\Sigma$ of ${\bf C}^3$ with coordinates $t,z,w$. 
Equivalently
we may embed the M-fivebrane in spacetime by a function of the 
form $F(t,z,w)=0$. Our first task then is to generalise the construction in
\cite{witten} and identify the appropriate complex
surface for \p{hyper} and its moduli.

We are interested here in manifolds $\Sigma$ which correspond
to two NS-fivebranes in the $x^0,x^1,x^2,x^3,x^4,x^5$ plane in ten dimensions. 
In this case we need two possible values of $t$ for every
point $(z,w)$, i.e. we need a two-sheeted cover of the $(z,w)$ plane. Thus
we assume that $F(t,z,w)$ takes the form
\be
A(z,w)t^2 - 2B(z,w)t + C(z,w) =0\ .
\ee
We do not want to consider here a configuration where there are semi-infinite
D-fourbranes coming off the two NS-fivebranes. To ensure this 
we must set  $A$ and $C$ to constants, so that $t$ is neither zero nor
infinite (corresponding to $s  =\pm\infty$) for any finite value of $z$ and 
$w$. 
Without loss of generality this leads to
$A=1$ and $C=\Lambda$. By rescaling $t$ and $B$ we can also
set $\Lambda=1$.
To obtain $N_1$ D-fourbranes in the $(x^1,w,x^6)$ plane 
and $N_2$ D-fourbranes in the $(x^1,z,x^6)$ plane we must chose $B$ 
to be a polynomial of degree $N_1$ in $z$ and
degree $N_2$ in $w$. The most general form for $B$ can be written as
\bea
B(z,w) &=& \alpha z^{N_1}w^{N_2} - p(w)z^{N_1} - q(z)w^{N_2}\nn
&&- \beta z^{N_1-1}w^{N_2-1}- r_1(w)z^{N_1-1}-r_2(z)w^{N_2-1}\nn
&&+ R(z,w)\ .\nn
\eea
Here $\alpha$ and $\beta$ are constants, $p$ and $r_1$ are polynomials of 
degree 
$N_2-1$ and $N_2-2$ in $w$ respectively,
$q$ and $r_2$ are polynomials of degree $N_1-1$ and $N_1-2$ in $z$ 
respectively and $R(z,w)$ is a 
polynomial of degree $N_1-2$ in $z$ and $N_2-2$ in $w$. 

We can set $\alpha=1$ by rescaling $z$ and $w$. This leaves $(N_1+1)(N_2+1)-1$
coefficients in $B$ to determine. 
To this end we imagine that $t$ is fixed. This
gives us a surface in the $z,w$ plane representing the two types of intersecting
D-fourbranes. For large $z$ with $w$ fixed this surface takes the form
\be
z(w^{N_2} - p(w)) - (q_{N_1-1}w^{N_2} + bw^{N_2-1} + r_1(w)) = 0\ ,
\label{bigz}
\ee
where  $q_{N_1-1}$ is the leading
coefficient in $q(z)$ and we have ignored lower order terms. 
To describe the intersecting brane configuration
we require that for large $z$ we go to a fixed value of $w$, representing
a D-fourbrane in the $(x^1,z,x^6)$ plane. In addition there should be a
symmetry between $z\rightarrow\infty$ and $z\rightarrow -\infty$.
This requires that both terms in \p{bigz} vanish separately. In particular we 
must be at a zero of $w^{N_2}-p(w)$ and if we write
\be
w^{N_2} - p(w) = \prod_{i'=1}^{N_2}(w - w_{i'})\ ,
\ee
then we can interpret the $w_{i'}$ as the locations of the $N_2$ D-fourbranes
in the $(x^1,z,x^6)$ plane. We can also shift the $w$ coordinate so 
that the centre of mass
is frozen at $w=0$, thereby setting $\sum_{i'}w_{i'}=0$. This means
that $p(w)$ is in fact only of degree $N_2-2$.
Similarly, by examining the large $w$ limit,  we write
\be
z^{N_1} - q(z) = \prod_{i=1}^{N_1}(z - z_i)\ ,
\ee
and interpret the $z_i$ as the locations of the $N_1$ D-fourbranes in the
$(x^1,w,x^6)$ plane. By a shift in the $z$ coordinate we can set the
centre of mass to $z=0$. Again this means that $\sum_{i}z_{i}=0$ and 
$q(z)$ is only of degree $N_1-2$.

It also follows from \p{bigz} that the polynomial 
(recall that now $q_{N_1-1}=0$)
\be
\beta w^{N_2-1} + r_1(w)\ ,
\ee
has $N_2$ roots precisely at the points $w_{i'}$. But 
this is a polynomial of degree
$N_2-1$ and therefore it can only
have $N_2$ roots if it is identically zero.
Hence we must have $\beta=r_1(w)=0$. Similarly from the
large $w$ analysis we find that $r_2(z) = 0$.
Thus $B$ can be written as
\be
B(z,w) = z^{N_1}w^{N_2} - p(w)z^{N_1} - q(z)w^{N_2} + R(z,w)\ ,
\ee
where now $p(w)$ is of degree $N_2-2$ and $q(z)$ is of degree $N_1-2$. 

Thus we find that there are $(N_1-1)+(N_2-1)$ moduli
coming from the polynomials $p$ and $q$ and also $(N_1-1)(N_2-1)$ 
moduli from the polynomial $R$ which we write in the form
\be
R(z,w) = \sum_{i=1}^{N_1-1}\sum_{i'=1}^{N_2-1} u_{ii'}z^{i-1}w^{i'-1}\ .
\ee
We have seen above that we can identify the moduli $z_i$ and
$w_{i'}$ with the  positions of the D-fourbranes. However 
the moduli in $R$ determine subtle features
of the hypersurface $\Sigma$ which are hard to see in the large $z,w$ limit
where we can clearly identify distinct D-fourbranes. 
It follows from
M-theory/type IIA duality that we can associate to the D-brane 
Yang-Mills theory
the complex hyper-surface in ${\bf C}^3$  given by
\be
y^2 = \left[z^{N_1}\prod_{i'}(w-w_{i'}) 
+ w^{N_2}\prod_i(z-z_i) - z^{N_1}w^{N_2} 
+ \sum_{ii'} u_{ii'}z^{i-1}w^{i'-1} \right]^2 - 1\ ,
\label{surface}
\ee
where $y=t-B$. Clearly for $w$ ($z$) fixed to a constant this surface 
reduces to the Seiberg-Witten curve for $N=2$ $SU(N_1)$ ($SU(N_2)$) 
gauge theory after a trivial rescaling.

It will be helpful to re-derive this surface by considering only features of 
the two four-dimensional
$SU(N)$ gauge theories. From this point of view we want to
construct a complex surface of the form
\be
y^2 = B^2 - 1\ ,
\ee
with $B$ a polynomial of degree $N_1$ in $z$ and $N_2$ in $w$. 
Furthermore we require that the
coefficients in $B$ are constructed from scalar 
gauge invariant operators of the four-dimensional Yang-Mills theories. 
It is easy to see
that for the  groups $SU(N_1)$ and $SU(N_2)$ 
the only independent such  operators are 
($i = 1,...,N_1-1\ ,\  i' = 1,...,N_2-1)$
\bea
\tilde z_i &=& {\rm Tr} (a^{N_1 +1-i})\ ,\nn
\tilde w_{i'} &=& {\rm Tr} (b^{N_2+1-i'})\ ,\nn
\tilde u_{ii'} &=& {\rm Tr} (a^{N_1+1-i}b^{N_2+1-i'})\ .\nn
\label{operators}
\eea

In the brane configuration \p{config} the $SO(1,9)$ Lorentz group is
broken down to $SO(1,1)\times SO(3)\times SO(2)\times SO(2)$
corresponding to the two-dimensional Lorentz group and rotations in the
$(x^7,x^8,x^9)$, $(x^4,x^5)$ and $(x^2,x^3)$ planes respectively. These
last three transformations appear in the gauge theory as an 
$SU(2)\times U(1)\times U(1)$ R-symmetry. If we consider the four-dimensional 
$SU(N)$ gauge theory on each type of D-fourbrane (and ignore the presence of 
the other type of D-fourbrane) then the corresponding
$U(1)$ R-symmetry is broken by quantum effects to ${\bf Z}_{2N}$. 
This group is generated by
\be
a \rightarrow e^{i\pi\over N_1}a\ ,\quad\quad 
b \rightarrow e^{i\pi\over N_2}b\ .
\ee
Thus if we denote the weights of a field under 
${\bf Z}_{2N_1}\times {\bf Z}_{2N_2}$ by $(p,q)$ then the operators transform
as
\bea
\tilde z_i &=& (N_1+1-i, 0)\ ,\nn
\tilde w_{i'} &=&(0, N_2+1-i')\ ,\nn
\tilde u_{ii'} &=&(N_1+1-i,N_2+1-i')\ .\nn
\eea
We require that ${\bf Z}_{2N_1}\times {\bf Z}_{2N_2}$ must be a 
symmetry of the surface. Let us  write 
\be
B =  \sum_{k=0}^{N_1}\sum_{l=0}^{N_2} c_{k,l}z^{k}w^{l}
\ee
and  normalise the leading term $c_{N_1,N_2}=1$. Noting that 
$B\rightarrow-B$ is a symmetry of the surface we see that we can assign the
variables $z$ and $w$ the weights $(1,0)$ and $(0,1)$ respectively under
${\bf Z}_{2N_1}\times {\bf Z}_{2N_2}$. 
Now consider an arbitrary term  in $B$. Invariance of $B$
(up to a minus sign) then tells us that $c_{k,l}$ must have weight
$(N_1-k,N_2-l)$. Thus we identify
\bea
c_{i-1,N_2} &=& \tilde z_{i}\ ,\nn
c_{N_1,i'-1} &=& \tilde w_{i'}\ ,\nn
c_{i-1,i'-1} &=& \tilde u_{ii'}\ .
\eea
However the coefficients of the sub-leading terms $c_{N_1-1,l}$ and 
$c_{k,N_2-1}$ have weights $(1,N_2-l)$ and $(N_1-k,1)$ respectively.
Since there are no gauge invariant operators with these dimensions we
must set $c_{N_1-1,l}=c_{k,N_2-1}=0$. Thus we find
\be
B = z^{N_1}w^{N_2} + \sum_{i} \tilde z_i z^{i-1}w^{N_2} 
+ \sum_{i'}\tilde w_{i'}z^{N_1}w^{i'-1}
+\sum_{ii'}\tilde u_{ii'}z^{i-1}w^{i'-1}\ .
\ee
This is precisely the same surface that we obtained from the geometry
of the intersecting branes with $\tilde u_{ii'} = u_{ii'}$,  
$\tilde z_i$  identified 
with the symmetric polynomial of degree $N_1+1-i$
in $z_i$ and $\tilde w_{i'}$ identified 
with the symmetric polynomial of degree $N_2+1-i'$ in $w_{i'}$. 

Just as was the case for the $N=2$ models this picture also suggests that
the moduli of surface are the scalar modes in the low energy effective 
description of the Yang-Mills theory. 
However one can easily see that not all of these scalar modes are dynamical
in the low energy M-fivebrane theory. To see this  consider the
dynamics of the scalar modes alone (i.e. ignore for now the three-form field
$H$). In this case the M-fivebrane dynamics 
are given by the standard $p$-brane action
\be
S = \int d^6x \sqrt{-{\rm det}g}\ ,
\ee
where $g_{mn}$ is the induced metric
\be
g_{mn} = \eta_{mn} + {1\over 2}\partial_m s\partial_n \bar s
+ {1\over 2}\partial_m \bar s\partial_n s \ .
\ee
Expanding this expression out to second order in two-dimensional derivatives
gives, ignoring a surface term, 
\be
S = {1\over2}\int d^2x d^2 w d^2 z \partial_\mu s\partial^{\mu}\bar s\ .
\label{action}
\ee
To evaluate the low energy scalar dynamics we promote the moduli of the surface
to two-dimensional scalar fields so that
\be
\partial_{\mu} s = \sum_i \partial_{\mu}z_i{\partial s\over \partial z_i}
+\sum_{i'}  \partial_{\mu}w_{i'}{\partial s\over \partial w_{i'}}
+ \sum_{ii'} \partial_{\mu}u_{ii'}{\partial s\over \partial u_{ii'}}\ .
\ee
Substituting this into the action $S$ one finds non-convergent integrals
(in the large $z,w$ limit) coming from the kinetic terms for the 
$z_i$ and $w_{i'}$ moduli. Thus we are forced to set 
the positions of the D-fourbranes $z_i(x)$ and $w_{i'}(x)$ to  
constants in the low energy theory. This is not unexpected since
the D-fourbranes are infinitely heavy from the point of view of the 
two-dimensional intersection. 

From the alternative derivation of the surface \p{surface} above we also see 
that 
we must identify the moduli with the vacuum expectation values of the 
gauge invariant operators \p{operators}. In particular the $z_i$ and 
$w_{i'}$ are
the $N_1-1$ and $N_2-1$ components of $a$ and $b$ in the Cartan subalgebra of 
$SU(N_1)$ and $SU(N_2)$ respectively. 
Using supersymmetry it follows that the superpartners of these
fields, namely $\chi_A,\chi_B,A_{\mu},B_{\mu}$, 
are also  non-dynamical.
This leaves us with $(N_1-1)(N_2-1)$ complex moduli $u_{ii'}$ and their
superpartners which we expect
to have smooth finite energy behaviour at low energy. From the above
analysis we see that these are identified with the composite 
gauge theory operators 
\be
u _{ii'} = {\rm Tr}(a^{N_1+1-i}b^{N_2+1-i'})\ .
\ee

\section{The M-Fivebrane Equations and Soliton Dynamics}

Before proceeding to analyse the dynamics of the configuration \p{hyper} 
we must first discuss the
equations of motion for the the M-fivebrane in flat eleven-dimensional 
spacetime. We then use these to find the  equations of motion
for the low energy motion of our string soliton solution on the 
M-fivebrane worldvolume.
We will use the equations
of motion found in \cite{hsw} from the superembedding approach
applied to the M-fivebrane \cite{hs}. 

The bosonic fields of the M-fivebrane consist of a closed three-form
$H_{mnp}$ and five scalars $X^{a'}$, $a'=6,7,8,9,10$, representing the
transverse fluctuations. We denote the worldvolume coordinates by $x^m$, 
$m=0,1,2,..,5$ and tangent frame indices by $a,b,c = 0,1,2,...,5$. The
two-dimensional coordinates $x^0,x^1$ are denoted by $x^{\mu}$.
The complex coordinates $z,w$ and $\bar z,\bar w$ are denoted by
$z^\alpha$ and $\bar z^{\bar \alpha}$ respectively.
The scalar fields define a worldvolume metric and vielbein
\bea
g_{mn} &=& \eta_{mn} + \partial_mX^{a'}\partial_n X^{a'}\ ,\nn
&=&\eta_{ab}e_m^{\ a}e_n^{\ b}\ , \nn
\eea
obtained through the pull-back of the flat eleven-dimensional metric.

In the full non-linear theory the three-form
$H_{mnp}$ satisfies a complicated self-duality constraint. This is obtained
by first considering a (linearly) self-dual three-form $h_{mnp}$
\be
h_{mnp} = {1\over 3!}\epsilon_{mnpqrs}h^{qrs}\ ,
\label{selfdual}
\ee
which is not assumed to be closed. We can then construct the tensor
\be
m_a^{\ b} = \delta_a^{\ b} - 2h_{acd}h^{bcd}\ ,
\ee
and define
\be
H_{mnp} = e_m^{\ a}e_n^{\ b}e_p^{\ c}(m^{-1})_c^{\ d}h_{abd}\ .
\ee
Thus $H_{mnp}$ satisfies a non-linear self-duality constraint
inherited from the (linear) self-duality of $h_{mnp}$.

We may now write the equations of motion for the bosonic fields  in the form
\cite{hsw}
\bea
G^{mn}\nabla_{m}\nabla_{n}X^{a'}&=&0\ ,\nn
G^{mn}\nabla_{m}H_{npq}&=&0\ ,\nn
\label{eqofm}
\eea
where 
\be
G^{mn} = e_a^{\ m}e_b^{\ n}m_c^{\ a}m_d^{\ b}\eta^{cd} \ ,
\ee
is a second metric incorporating the three-form field. If $H_{mnp}$ 
is constructed
as above and hence  satisfies its non-linear self-duality constraint then the
second equation in \p{eqofm} is equivalent to the closure of 
$H_{mnp}$ \cite{hsw}.

To find the low energy motion of the string soliton it is sufficient to
consider only terms up to second order in spacetime derivatives 
$\partial_\mu$
and the three-form $H_{mnp}$. In this case 
we may simply write  
\be
H_{mnp} = e_m^{\ a}e_n^{\ b}e_p^{\ c}h_{abc}\ .
\label{Hh}
\ee
It is  understood in \p{Hh} 
that $H$ appears in the worldvolume frame and $h$ in the tangent frame.

Once the self-duality condition on $H$ is satisfied
the low energy equations of motion for the M-fivebrane are found from \p{eqofm}
to be
\be
\partial_\mu\partial^\mu s - \partial_\alpha\left[
{\partial_\mu s\partial^\mu s\bar \partial_{\bar \alpha}\bar s \over
1 + |\partial s|^2}\right] 
- 4 H_{mn}^{\ \ \ \alpha}H^{mn\beta}\partial_\alpha\partial_\beta s = 0
\ ,
\label{sreduc}
\ee
for the scalar fields, where 
$|\partial s|^2 =|\partial_z s|^2+|\partial_w s|^2$. 
For the three-form  we may simply use
\be
\partial_{[m}H_{npq]}=0\ .
\ee
Here we have assumed that only the two scalars 
$X^6$ and $X^{10}$ are non-constant.
In particular the closure of $H_{mnp}$ gives several equations
\bea
\partial_{[\mu}H_{\nu \alpha\beta]}&=&0\ ,
\nonumber\\
\partial_{[\mu}H_{\nu \alpha\bar\beta]}&=&0\ ,
\nonumber\\
\partial_{[\mu}H_{\alpha\beta\bar\gamma]}&=&0\ ,\nonumber \\
\partial_{[\alpha}H_{\beta\bar\gamma\bar\delta]}&=&0\ ,\nonumber \\
\label{Heqns}
\eea
and their complex conjugates.

To obtain the low energy motion for the soliton
we let its moduli depend on $x^\mu$ and substitute the form for $s(z,w)$
into these equations. We must then dimensionally reduce the scalar 
equations to two-dimensions. Following the methods of \cite{lw} we consider
\be
\int_\Sigma \left\{\partial_\mu\partial^\mu s - \partial_\alpha\left[
{\partial_\mu s\partial^\mu s\bar \partial_{\bar \alpha}\bar s \over
1 + |\partial s|^2}\right] 
- 4 H_{mn}^{\ \ \ \alpha}H^{mn\beta}\partial_\alpha\partial_\beta s 
\right\}dz \wedge dw \wedge \theta^I =0 \ ,
\ee
where the $\theta^I$ are a basis for $H^2(\Sigma)$. 
For the three-form we consider only those equations which are second order 
in spacetime  indices $\mu$ to be equations of motion. For these we consider
\bea
\int_\Sigma \partial_{[\mu}H_{\nu \alpha\beta]}dz^{\alpha}\wedge 
dz^{\beta}\wedge \theta^I =0\ ,\nonumber\\
\int_\Sigma \partial_{[\mu}H_{\nu \alpha\bar\beta]}
dz^{\alpha}\wedge dz^{\bar\beta}\wedge 
\theta^I =0\ .\nonumber\\
\label{Hreduc}
\eea
This leaves us with the last two equations in \p{Heqns} which we view 
as constraints (along with self-duality) on  $h_{mnp}$. 

It is helpful at this point 
to list the non-vanishing components of the vielbein
\bea
e_{\alpha}^{\ \beta} &=& \delta_{\alpha}^{\ \beta}
+\left({\sqrt{1+|\partial s|^2}-1\over |\partial s|^2}\right)
\partial_\alpha s\bar\partial_{\bar\beta}\bar s\ 
+ {\cal O}((\partial_\mu )^2),\nn
e_\mu^{\ \alpha} &=& {\partial_\mu s\bar \partial_{\bar\alpha} \bar s
\over\sqrt{1+|\partial s|^2}}
+{\cal O}((\partial_\mu )^3)\ ,\nn
e_\mu^{\ \nu} &=& \delta_\mu^\nu\ + {\cal O}((\partial_\mu )^2),\nn
\eea 
plus their complex conjugates. 
In this paper we will only consider ansatze for the three-form which satisfy
\bea
H_{\mu\nu\alpha}= 0\ .
\eea
From \p{Hh} we find that
\be
h_{\mu\nu\alpha} = {\cal O}((\partial_{\mu})^2)\ ,
\ee
and it follows from \p{selfdual} that 
$h_{\alpha\beta\bar\gamma} = {\cal O}((\partial_{\mu})^2)$. Thus we
simply find 
\be
H_{\mu\alpha\beta} = 
e_{\alpha}^{\ \gamma}e_{\beta}^{\ \delta}h_{\mu\gamma\delta}\ ,\quad 
H_{\mu\alpha\bar\beta} =e_{\alpha}^{\ \gamma}e_{\bar\beta}^{\ \bar\delta}
h_{\mu\gamma\bar\delta}\ ,
\ee
where $H$ is in the world frame and $h$ is in the tangent frame.
We may now consider 
ansatze for $H_{\mu\alpha\beta}$ and $H_{\mu\alpha\bar\beta}$
of the form
\be
H_{\mu\alpha\beta} = {\cal U}_{\mu}\kappa_{\alpha\beta}\  ,\quad
H_{\mu\alpha\bar\beta} = 
{\cal V}^+_{\mu}\kappa^+_{\alpha\bar\beta}
+{\cal V}^-_{\mu}\kappa^-_{\alpha\bar\beta}\  ,
\label{Han}
\ee
where we have chosen $\kappa^\pm_{\alpha\bar\beta}dz^\alpha\wedge 
d{\bar z}^{\bar\beta}$ to be (anti-) self-dual forms on $\Sigma$.
Substituting this into the self-duality condition we learn that 
${\cal U}_{\mu} = \epsilon_{\mu\nu}{\cal U}^{\nu}$ and  
${\cal V}^\pm_{\mu}=\pm \epsilon_{\mu\nu}{\cal V}^{\pm\nu} $ are  
two-dimensional chiral fields.

If we assume that the three fields
${\cal U}_\mu$ and ${\cal V}_\mu^\pm$ are independent then the  third 
equation in \p{Heqns}  implies that 
$\kappa_{\alpha\beta}dz^\alpha\wedge dz^\beta$ and 
$\kappa^\pm_{\alpha\bar\beta}dz^\alpha\wedge d{\bar z}^{\bar\beta}$ 
are closed $(2,0)$ and $(1,1)$ forms on $\Sigma$ respectively.
It is not hard to see that a non-zero $H_{\mu\nu\alpha}$ would correspond
to $(1,0)$ forms on $\Sigma$.
Finally we must consider the fourth equation in \p{Heqns}. We expect that
this equation leads to no additional constraints on $H$. 
The reason for this is that if we
integrate it  over the manifold $\Sigma$ we obtain the constraint
\be
\int_\Sigma\partial_{[\alpha}H_{\beta\bar\gamma\bar\delta]}
dz^{\alpha} \wedge dz^{\beta} \wedge dz^{\bar\gamma} \wedge dz^{\bar\delta}
= \int_\Sigma dH = 0\ ,
\ee
which is identically true if $\Sigma$ has no boundary or 
if the fields which vanish at infinity.
Thus the ansatze for chiral the fields on a general $\Sigma$ takes the form
\be
H_{\mu\alpha\beta} 
= \sum_{I=1}^{b^{(2,0)}} {\cal U}_{\mu I}\omega^I_{\alpha\beta}\ ,\quad
H_{\mu\alpha\bar\beta}
=\sum_{I=1}^{b^{+(1,1)}}{\cal V}^+_{\mu I}\omega^{+I}_{\alpha\bar\beta}
+\sum_{I=1}^{b^{-(1,1)}}{\cal V}^-_{\mu I}\omega^{-I}_{\alpha\bar\beta}\ .
\label{Hans}
\ee
Here $\omega^I$, $\omega^{+I}$ and $\omega^{-I}$ form  basis of
$H^{(2,0)}(\Sigma)$, $H^{+(1,1)}(\Sigma)$ and  $H^{-(1,1)}(\Sigma)$ 
respectively, where $H^{\pm(1,1)}(\Sigma)$ is the  
(anti-) self-dual subgroup of $H^{(1,1)}(\Sigma)$. 
The index $I$ is understood to run 
over the dimension of the appropriate cohomology group. 
The first two equations in \p{Heqns} now give the following three equations
for ${\cal U}_{\mu I}$, ${\cal V}^+_{\mu I}$ and ${\cal V}^-_{\mu I}$
\bea
0&=& \partial_{[\mu}{\cal U}_{\nu]J} \omega^J
+\partial_{[\mu}u^A{\cal U}_{\nu]J} 
{\partial\omega^J\over \partial u^A}\ ,\nn
0&=& \partial_{[\mu}{\cal V}^+_{\nu]J}\omega^{+J}
+\partial_{[\mu}u^A{\cal V}^+_{\nu]J}
{\partial\omega^{+J}\over \partial u^A}\ ,\nn
0&=& \partial_{[\mu}{\cal V}^-_{\nu]J} \omega^{-J}
+\partial_{[\mu}u^A{\cal V}^-_{\nu]J} 
{\partial\omega^{-J}\over \partial u^A}\ .\nn
\eea
With these ansatze  for the chiral fields we also find the full
scalar equation
\bea
0&=&\partial_\mu\partial^\mu u^A
{\partial s\over\partial u^A}
+\partial_\mu u^A\partial^\mu u^B
{\partial^2 s\over \partial u^A \partial u^B}\nn
&&- \partial_\mu u^A\partial^\mu u^B\partial_\alpha\left[
{\partial s\over\partial u^A}{\partial s\over \partial u^B}
{\bar \partial_{\bar \alpha}\bar s\over 1+|\partial s|^2}\right]
\nn
&&-4{\bar{\cal U}}_{\mu J}{\cal V}_K^{-\mu}
\left(g^{\alpha\bar\gamma}g^{\beta\bar\delta}g^{\epsilon\bar\eta}
\bar\omega^J_{\bar\eta\bar\delta}
\omega^{-K}_{\epsilon\bar\gamma}
\partial_\alpha\partial_\beta s\right)
\ .
\label{scalareqn}
\eea

Finally we may now write
the two-dimensional equations of motions. 
For the scalars we find
\bea
0&=&\partial_\mu\partial^\mu u^A 
\int_{\Sigma}{\partial s\over\partial u^A}dz\wedge dw\wedge \theta^I
+\partial_\mu u^A\partial^\mu u^B
\int_{\Sigma}{\partial^2 s\over \partial u^A \partial u^B}
dz\wedge dw\wedge \theta^I\nn
&&- \partial_\mu u^A\partial^\mu u^B\int_\Sigma\partial_\alpha\left[
{\partial s\over\partial u^A}{\partial s\over \partial u^B}
{\bar \partial_{\bar \alpha}\bar s\over 1+|\partial s|^2}\right]
dz\wedge dw \wedge\theta^I\nn
&&-4{\bar{\cal U}}_{\mu J}{\cal V}_K^{-\mu}\int_\Sigma
\left(g^{\alpha\bar\gamma}g^{\beta\bar\delta}g^{\epsilon\bar\eta}
\bar\omega^J_{\bar\eta\bar\delta}
\omega^{-K}_{\epsilon\bar\gamma}
\partial_\alpha\partial_\beta s\right)
dz\wedge dw\wedge\theta^I\ ,
\label{scalarequation}
\eea
where $u^A$ are the moduli for the surface $s(z,w)$. It is easy to see
that only those $\theta^I$ in $H^{(0,2)}(\Sigma)$ give a non-vanishing 
contribution to the integrals in \p{scalarequation}. We also note that if
if ${\cal U}_\mu={\cal V}^\pm_\mu=0$ then the equations of motion for
the scalars $u^A$ can be derived from the action \p{action}. However this
only leads to the first two terms in \p{scalarequation}. Thus we expect that
the third integral in \p{scalarequation} always vanishes. 
This is not surprising
since, for $(0,2)$ forms $\theta^I$, the integrand is a
total derivative. 
We also find the following equations for the chiral fields
\bea
0&=& \partial_{[\mu}{\cal U}_{\nu]J}\int_\Sigma \omega^J\wedge\theta^I
+\partial_{[\mu}u^A{\cal U}_{\nu]J}\int_\Sigma 
{\partial\omega^J\over \partial u^A}\wedge\theta^I\ ,\nn
0&=& \partial_{[\mu}{\cal V}^+_{\nu]J}\int_\Sigma \omega^{+J}\wedge\theta^I
+\partial_{[\mu}u^A{\cal V}^+_{\nu]J}\int_\Sigma 
{\partial\omega^{+J}\over \partial u^A}\wedge\theta^I\ ,\nn
0&=& \partial_{[\mu}{\cal V}^-_{\nu]J}\int_\Sigma \omega^{-J}\wedge\theta^I
+\partial_{[\mu}u^A{\cal V}^-_{\nu]J}\int_\Sigma 
{\partial\omega^{-J}\over \partial u^A}\wedge\theta^I\ .\nn
\label{chiralequations}
\eea
Since the wedge product of a self-dual and an anti-self-dual form vanishes,
it follows that the only non-vanishing
integrals in the equations for ${\cal U}_\mu$, ${\cal V}_\mu^+$ and 
${\cal V}_\mu^-$ come from $\theta^I$ in $H^{(0,2)}(\Sigma)$, 
$H^{+(1,1)}(\Sigma)$ and  $H^{-(1,1)}(\Sigma)$ respectively.

\section{Effective Dynamics of Two-Dimensional Yang-Mills Theories}

In this section we will explicitly derive the low energy effective dynamics
for an M-fivebrane wrapped around a four-cycle given by \p{surface}. This can
then be interpreted as the M-theory prediction for the low energy effective
theory of the two-dimensional $(4,0)$ Yang-Mills theory.

Before we proceed we must construct a suitable ansatz for $H_{mnp}$. A
non-zero $H_{\mu\nu z}$ would lead to vector modes in the low energy
dynamics and since these contain no propagating degrees of freedom we
set them to zero.
In addition, in 
this section it will be sufficient to consider $H_{\mu\alpha\bar\beta}=0$, 
which we will justify below. Thus we are only concerned with the first
equation in \p{Han} corresponding to $(2,0)$ forms on $\Sigma$.
It is not hard to check that a basis of normalisable $(2,0)$ forms is 
given by
\be
\omega^{ii'} \equiv {z^{i-1}w^{i'-1}\over y} dz\wedge dw 
= -{1\over R}{\partial s\over \partial u_{ii'}} dz\wedge dw \ ,
\ee
where $y$ is given by \p{surface}. Thus the general form for 
$H_{\mu\alpha\beta}$ is
\be
H_{\mu\alpha\beta} = \sum_{ii'}{\cal U}_{\mu ii'}\omega^{ii'}_{\alpha\beta}\ .
\ee

Since we have set $H_{\mu\nu\alpha}=H_{\mu\alpha\bar\beta}=0$ one finds that
the scalar equation of motion simplifies to 
\be
\partial_\mu\partial^\mu s - \partial_\alpha\left[
{\partial_\mu s\partial^\mu s\bar \partial_{\bar \alpha}\bar s \over
1 + |\partial s|^2}\right] =0\ .
\label{ttt}
\ee
Performing the  reduction of the equations of motion
given in the last section yields
\be
\partial_{\mu}\partial^{\mu} u_{ii'}I^{ii'\ jj'} + 
\partial_{\mu}u_{ii'}\partial^{\mu}u_{kk'}
{\partial I^{ii'\ jj'}\over \partial u_{kk'}} + 
\partial_{\mu}u_{ii'}\partial^{\mu}u_{kk'}K^{ii'\ kk'\ jj'}=0\ ,
\ee 
where 
\be
I^{ii'\ jj'} = \int_\Sigma \omega^{ii'}\wedge \bar\omega^{jj'}\ ,
\ee
and
\be
K^{ii'\ kk'\ jj'} = \int_\Sigma\partial_\alpha\left[
{\omega^{ii'}_{zw}\omega^{kk'}_{zw}\bar \partial_{\bar \alpha}\bar s \over
1 + |\partial s|^2}\right]dz\wedge dw \wedge \bar\omega^{jj'}\ . 
\ee
Since the integrand is a total derivative, and
the singularities are mild, it is not hard to see that $K^{ii'\ kk'\ jj'}=0$, 
as we expected in the previous section.
Thus we find the equations of motion
\be
\partial_{\mu}\partial^{\mu} u_{ii'}I^{ii'\ jj'} + 
\partial_{\mu}u_{ii'}\partial^{\mu}u_{kk'}
{\partial I^{ii'\ jj'}\over \partial u_{kk'}}\ ,
\ee
for the $(N_1-1)(N_2-1)$ complex scalars $u_{ii'}$. 
Indeed the above equations of motion are then precisely those of the
action obtain by  dimensionally reducing  \p{action}. In this way we find
the the two-dimensional action
\be
S = \int
d^2x\partial_{\mu}u_{ii'}\partial^{\mu}{\bar u}_{jj'}I^{ii'\ jj'}\ ,
\ee
for the scalars $u_{ii'}$.

Next we must consider the equation of motion for $H_{\mu\alpha\beta}$.
This is obtained using \p{chiralequations}  to give
\be
\partial_{[\mu}{\cal U}_{\nu] ii'}I^{ii'\ jj'} 
+ \partial_{[\mu}u_{kk'}{\cal U}_{\nu]ii'}
{\partial I^{ii'\ jj'}\over \partial u_{kk'}} = 0\ .
\ee
Here we find chiral equations of motion for $(N_1-1)(N_2-1)$ complex bosons.
We have yet to account for the  zero modes of
$H_{\mu\alpha\bar\beta}$.  These are obtained 
from $(1,1)$-forms on $\Sigma$. 
However, since we have accounted for all of the low energy
fields of the D-fourbranes, these fields, if any exist, are not 
zero modes of our worldvolume string soliton and so we are justified in
ignoring them.

Before ending this section we will make some observations about the
effective action that we have derived. 
For concreteness let us consider the simplest case where
$N_1=N_2=2$, i.e. for the group $SU(2)\times SU(2)$. We therefore have 
only one complex scalar mode $u$ and the surface \p{surface} takes the form
\be
y^2 = \left[z^2w^2-w_0^2z^2 - z_0^2w^2+u \right]^2 - 1\ .
\ee
The equations of motion for the fields are now 
\bea
\partial_{\mu}\partial^{\mu} u I + 
\partial_{\mu}u\partial^{\mu}u
{\partial I\over \partial u}=0\ ,\nn
\partial_{[\mu}{\cal U}_{\nu]}I 
+ \partial_{[\mu}u{\cal U }_{\nu]}
{\partial I\over \partial u} = 0\ ,\nn
\eea
with
\be
I = \int_\Sigma \omega \wedge\bar\omega
= \int {d^2 w d^2 z\over y\bar y} \ .
\ee
One can show by a straightforward change of variables that $I$ can be
written as
\be
I = \int {d^2\xi \over |\xi^2 - z_0^2w_0^2|} 
I_{SW}(\xi^2-u) \ ,
\ee
where 
\be
I_{SW}(x) \equiv \int d^2\zeta {1\over |(\zeta^2- x)^2 - 1|} \ ,
\ee
is the standard four-dimensional Seiberg-Witten elliptic integral for $SU(2)$
(in the notation of \cite{sw}, $I_{SW}(u)=|da/du|^2{\rm Im}\tau$). 
From the known form for this function \cite{sw} one sees that for large
$x$, $I_{SW}(x)\propto {\rm ln}x/|x|$ 
so that $I$ is convergent for large $\xi$. 
In addition one can see that the singular points of 
$I_{SW}(\xi^2-z_0^2w_0^2)$ are
only logarithmic. Therefore, even if they occur at a zero of 
$\xi^2-z_0^2w_0^2$, $I$ remains finite and well defined. 

Thus we obtain smooth low energy effective
dynamics. Furthermore all these quantities are holomorphic and so one is
strongly lead to believe that they are in fact the correct low energy
dynamics, including all the quantum corrections. Note also that there is
no value for $u$ where $I$ is singular. This is radically different 
to the solution of Seiberg and Witten \cite{sw} where there are points on the
$u$ plane where the low energy effective action is singular (i.e. points
where $F_{SW}$ has a logarithmic singularity).
As is well known these points are associated with  BPS states becoming 
massless. Thus we conclude here that there are no points on the $u$ plane
where extra light degrees of freedom need to be included. Note though that
$I$ is singular if $z_0=0$ or $w_0=0$. This corresponds to putting two of the 
parallel D-fourbranes on top of each other, leading to a restoration in the 
non-Abelian gauge symmetry.

\section{Black Hole Effective Field Theory}

Let us now consider a different form for the surface 
\p{hyper}.
In particular, consider M-theory compactified on ${\cal M}\times S^1$,
where ${\cal M}$ is a six-dimensional Calabi-Yau manifold. This results in
four-dimensional $N=1$ supergravity for the M-theory low energy effective
action. If we wrap the M-fivebrane on a four-cycle in $\cal M$ and then the
resulting string soliton over the $S^1$ we obtain a  black hole in
four dimensions, which is a BPS solution of the  supergravity theory.
The entropy of
this black hole was found  in
\cite{msw,vafa} by computing the number of degrees of freedom of the
black hole. The latter arise from the moduli of the embedded cycle,
the three-form field of the fivebrane, as well as the momenta on the
$S^1$. These are the same fields that we discussed in section four above. 
Therefore in 
this  section we will use the analysis described in section four to 
compute the  precise effective low energy
dynamics of the wrapped M-fivebrane. Just as for the previous case  we 
obtain the low energy dynamics by letting the
moduli of the M-fivebrane string soliton depend on the $x^\mu$ coordinates. 
Although we
do not carry out the remaining $S^1$ reduction, this is
straightforward and yields the low energy dynamics of the
back hole.  Again the resulting scalar degrees of freedom  posses $(4,0)$
supersymmetry. Effective actions for the conformal field theory of
black hole  internal degrees of freedom have also been constructed in
\cite{harstr,lambert,gt} in other contexts.

A key  difference with the work in the
previous section is that we now 
consider four-dimensional hypersurfaces $\Sigma$
of a six-dimensional complex space that can be compactified. 
In fact,  we will
consider any such  $\Sigma$ which has no $(1,0)$ or $(0,1)$ forms.
However
we will continue to use the M-fivebrane equations of
motion in a flat background. It is hoped that the resulting system of
equations provide a good description of an 
M-fivebrane wrapped on a four-cycle of a six-dimensional 
Calabi-Yau manifold. In principle it is straightforward to modify
our results by using the M-fivebrane equations of motion in a curved
background \cite{hsw}.

Since we consider $\Sigma$ to have  no one-cycles  we set again set
$H_{\mu\nu\alpha}=0$.
However now  we  need to consider the full
M-fivebrane dynamics, including zero modes of
$H_{\mu\alpha\bar\beta}$. Again we follow 
the discussion in section four.
The scalar fields that arise from
the moduli of $\Sigma$  obey the field equation arising from the
scalar field $s=X^6+iX^{10}$. In two-dimensions this is reduced to
\bea
0&=&\partial_\mu\partial^\mu u^A 
\int_{\Sigma}{\partial s\over\partial u^A}dz\wedge dw\wedge \theta^I
+\partial_\mu u^A\partial^\mu u^B
\int_{\Sigma}{\partial^2 s\over \partial u^A \partial u^B}
dz\wedge dw\wedge \theta^I\nn
&&-4{\bar{\cal U}}_{\mu J}{\cal V}_K^{-\mu}\int_\Sigma
\left(g^{\alpha\bar\gamma}g^{\beta\bar\delta}g^{\epsilon\bar\eta}
\bar\omega^J_{\bar\eta\bar\delta}
\omega^{-K}_{\epsilon\bar\gamma}
\partial_\alpha\partial_\beta s\right)
dz\wedge dw\wedge\theta^I\ ,
\label{sceq}
\eea
where we have dropped the total derivative term in
\p{scalarequation}.
As is clear from this equation, if there are anti-self-dual $(1,1)$
forms on
$\Sigma$ then we see  that
there will be quadratic terms in $H$ in the scalar equation.
The three-form field is again expanded as in \p{Hans} and we 
find the equations of motion
\bea
\partial_{[\mu}{\cal U}_{\nu]I}K^{IJ}
+ \partial_{[\mu}u^A{\cal U }_{\nu]I}
{\partial K^{IJ}\over \partial u^A} &=& 0\ ,\nn
\partial_{[\mu}{\cal V}^+_{\nu]I}J^{IJ}
+ \partial_{[\mu}u^A{\cal V}^+_{\nu]I}
{\partial J^{IJ}\over \partial u^A} &=& 0\ ,\nn
\partial_{[\mu}{\cal V}^-_{\nu]I}L^{IJ}
+ \partial_{[\mu}u^A{\cal V}^-_{\nu]I^-}
{\partial L^{IJ}\over \partial u^A} &=& 0\ ,\nn
\eea
where $u^A$ are the moduli of the surface (which are the scalars in the
two-dimensional theory) and
\be
K^{IJ} = \int_\Sigma\omega^I\wedge\omega^J\ ,\quad
J^{IJ} = \int_\Sigma\omega^{+I}\wedge\omega^{+J} \ ,\quad
L^{IJ} = \int_\Sigma\omega^{-I}\wedge\omega^{-J}\ .
\label{JKL}
\ee

The above equations for the three-form  field moduli
do not apparently depend on many of the 
detailed features  of the fivebrane dynamics. However,
this can not be said of the equation of motion for the $u^A$
as a result of the last term in \p{sceq}. These
introduce complicated non-holomorphic integrals 
which are difficult to evaluate in
general (although it seems likely that 
they can be evaluated along the lines used in
\cite{lwsuper}). Let us therefore concentrate here on the equations of
motion for the chiral fields.

The $J,K$ and $L$ integrals in \p{JKL} 
can in general be evaluated using a Riemann
Bilinear  relation of the form
\be
\int_\Sigma \omega\wedge\lambda
= (\Omega^{-1})^{IJ}\int_{A_I} \omega \int_{A_J}\lambda
\ ,
\ee
where $\Omega_{IJ}$ is the intersection form of $\Sigma$ and $A_I$ are
a  basis of homology 2-cycles,
$I,J= 1,2,...,b^2(\Sigma)$.

Since the space of two-forms in the second de Rham cohomology and the
space of two-cycles of $\Sigma$ are isomorphic, we can choose the two-cycles 
to
be dual to the two-forms inherited from the
underlying complex structure. Thus there exists a division 
of cycles into $C^I$, $\bar C^I$ and $C^I_\pm$, with the 
index range understood to depend on which cycles are referred to,  such that
\be
\int_{C^I}w^J=\delta_I^J,\ \int_{\bar C^I}\bar w^J=\delta_I^J
,\ \int_{C^{I}_\pm} w^{\pm J}=\delta_I^J
\label{cycles}
\ee
with all other integrals zero.
With this choice of basis the Riemann bilinear identity implies that
the integrals in equation \p{JKL} are just given by the inverse of the
intersection matrix, for example $K_{IJ}=(I^{-1})_{IJ}$ (note  that
the range of the indices varies so $J,K$ and $L$ are distinct).
We note that this choice of basis is not that usually used in the physics
literature. Often the intersection matrix is assumed to have a simple
form, which does not depend on the moduli of the surface, whereas 
the self-duality condition on the forms 
contains a moduli dependent matrix. The integral of
the forms over the two-cycles then leads to the period matrix of the
manifold.  Given an arbitrary set of forms and their self-duality
properties it is a matter of linear algebra to compute the
transformation on the cycles to find a basis that satisfies
\p{cycles}.

To illustrate this point, let us consider a Riemann surface of
genus one which has a canonical set of one cycles $A$ and $B$ with the
intersection matrix $\Omega = \left(\matrix {0 &1\cr -1 &0\cr}\right)$, 
and a holomorphic one-form
$\lambda$. We can always normalise
$\lambda$ such that
\be
\int _A \lambda=1\ ,\quad  \int _B \lambda=\tau
\ee
Changing to  the one cycles $C,\ \bar C$ by
\be
C= -{\bar\tau\over \tau -\bar \tau}A+ {1\over \tau -\bar \tau}B\ , \quad
\bar C= {\tau\over \tau -\bar \tau}A- {1\over \tau -\bar \tau}B
\ee
the integrals take the simple form
\be
\int _C \lambda=1\ ,\int _{C} \bar \lambda=0,\  \int _{\bar C} \lambda=0,\
\int _{\bar C}\bar  \lambda=1\ .
\ee
However, the  intersection matrix becomes
\be
\Omega= -{1\over (\tau-\bar \tau)}
\left(\matrix {0 &1\cr -1 &0\cr}\right)\ .
\ee

In light-cone coordinates,
$X^\pm=X^0 \pm X^1$, the self-duality conditions on
${\cal U}_{\mu I}$ and ${\cal V}^\pm_\mu$ are readily 
analysed. Indeed the only non-vanishing components of ${\cal U}_{\mu I}$ are
${\cal U}_{-I}$ and the only
non-vanishing components of
${\cal V}^\pm_{\nu I}$ are ${\cal V}^+_{- I}$ and ${\cal V}^-_{+ I}$.
The above equation of motions then take a very simple form 
with the above choice of cycles.
For example, the  equation for
${\cal U}_{\mu I}$ becomes
\be
\partial_{+}{\cal U}_{- K}
+ \partial_{+}u^A{\cal U }_{- I}
(I^{-1})^{IJ}{\partial I^{JK}\over \partial u^A} = 0\ ,
\ee
and there are similar equations for ${\cal V}^\pm_{\mp I}$.

Thus we find a system of equations which describe the two-dimensional
motion of scalar degrees of freedom. The $u^A$ scalars which arise
from the moduli of the four-cycle are neither left nor right moving, but the
scalars the arise from the three-form  field  are either left
handed or right handed, depending on the (anti-) 
self-duality of the two-cycle they
correspond to.
These equations describe the motion of the degrees of freedom of a
black string, but making a trivial reduction on $S^1$ we find
equations that describe the internal dynamics of
a black hole. 

Clearly, these equations are classically
superconformally invariant, but their $(4,0)$ supersymmetry
is not enough to 
ensure that they are superconformally invariant at the quantum
level due to anomalies (see for example \cite{hp}). 
However, since these equations  arise from an Abelian M-fivebrane 
which should
have a consistent quantum theory, we might expect these equations to be
superconformally invariant in the full quantum theory. This is also
expected since the near horizon limit has an $AdS^3$ 
structure \cite{maldacena}.
One can also enquire if the above system of dynamics is integrable.
Clearly,  their conformal invariance implies that they
possess an infinite number of conserved quantities, whose
currents  are polynomials and moments of  the energy momentum tensor.
However, this not necessarily sufficient for integrability, 
since they may not distinguish between all the physical states. 

Finally, since the fivebrane equations of motion are known in the presence of
a non-trivial background \cite{hsw}, one could repeat the derivation in the
presence of the background to find the low energy dynamics of the
black hole in the presence of other matter. In this system one could
study the interaction of external matter and a black hole in detail.

\bigskip
\noindent
{\bf Acknowledgements}

\noindent
We would like thank David Olive and Andrew Pressley for helpful discussions.

\bigskip
\noindent 
Note Added:

\noindent
While this paper was being written up,  \cite{mmt} appear on the hep-th 
archive,  which provides
a similar discussion to that given in section six.



\bigskip
\begin{thebibliography}{99}

\bibitem{witten}
E. Witten, Nucl. Phys. {\bf B500} (1997) 3, hep-th/9703166.

\bibitem{bbs} 
K. Becker, M. Becker and A. Strominger, 
Nucl. Phys. {\bf B456} (1995) 130, hep-th/9507158.

\bibitem{gp}
G. Gibbons and G. Papadopoulos, {\it Calibrations and Intersecting Branes}, 
hep-th/9803163.

\bibitem{glw}
J.P. Gauntlett, N.D. Lambert and P.C. West,
{\it Branes and Calibrated Geometries}, hep-th/9803216, to appear in 
Comm. Math. Phys.

\bibitem{sw}
N. Seiberg and E. Witten,
Nucl. Phys. {\bf B426} (1994) 19, hep-th/9407087.

\bibitem{hsw}
P.S. Howe, E. Sezgin and P.C. West, Phys. Lett. {\bf B399} (1997) 49, 
hep-th/9702008.

\bibitem{hs}
P.S. Howe and E. Sezgin, Phys. Lett. {\bf B394} (1997) 62, hep-th/9611008.

\bibitem{hlw}
P.S. Howe, N.D. Lambert and P.C. West, Phys. Lett. {\bf B418} (1998) 85, 
hep-th/9710034.

\bibitem{lw}
N.D. Lambert and P.C. West, 
Nucl. Phys. {\bf B524} (1998) 141, hep-th/9712040.

\bibitem{lwsuper}
N.D. Lambert and P.C. West, 
Phys. Lett. {\bf B424} (1998) 281, hep-th/9801104.

\bibitem{hh}
A. Hanany and K. Hori,
Nucl. Phys. {\bf B513} (1998) 119, hep-th/97007192.

\bibitem{brodie}
J.H. Brodie, 
Nucl. Phys. {\bf B517} (1998) 36, hep-th/9709228.

\bibitem{alishahiha}
M. Alishahiha,
Phys. Lett. {\bf B420} (1998) 51, hep-th/9710020;
Nucl. Phys. {\bf B528} (1998) 171, hep-th/9802151.

\bibitem{gu}
H. Garcia-Compean and A.M. Uranga,
{\it Brane Box Realization of Chiral Gauge Theories in Two
Dimensions}, hep-th/9806177. 

\bibitem{en1}
W. Lerche,
{\it Fayet-Iliopoulos Potentials from Four Folds}, JHEP 9711 {\bf 004} (1997),
hep-th/9709146.

\bibitem{en2}
P. Kaste, {\it On the Twisted Chiral Potential in 2D and the Analog of
Rigid Special Geometry for Four Folds}, JHEP 9906 {\bf 021} (1999),
hep-th/9904218.

\bibitem{maldacena}
J.M. Maldacena, Adv. Theor. Math. Phys. {\bf 2} (1998) 231, hep-th/9711200. 

\bibitem{msw}
J. Maldacena, A. Strominger and E. Witten,
J. High Energy Phys. {\bf 12} (1997) 002, hep-th/9711053. 

\bibitem{vafa}
C. Vafa,
Adv. Theor. Math. Phys. {\bf 2} (1998) 207, hep-th/9711067.

\bibitem{klmvw}
A. Klemm, W. Lerche, P. Mayr, C. Vafa and N. Warner, Nucl. Phys. {\bf B477} 
(1996) 746, hep-th/9604034.

\bibitem{bsv}
M. Bershadsky, V. Sadov and C. Vafa, 
Nucl. Phys. {\bf B463} (1996) 398, hep-th/9510225.

\bibitem{hw}
A. Hanany and E. Witten,
Nucl. Phys. {\bf B492} (1997) 152, hep-th/9611230.

\bibitem{lwnew}
N.D. Lambert and P.C. West, 
``D-branes in the Green-Schwarz Formalism'', hep-th/9905031.

\bibitem{harstr}
J.A. harvey and A. Strominger, Nucl. Phys. {\bf B449} (1995) 535, 
hep-th/9504047.

\bibitem{lambert}
N.D. Lambert, Nucl. Phys. {\bf B519} (1998) 214, hep-th/9707156.

\bibitem{gt}
G. Gibbons and P.K. Townsend, {\it Black Holes and Calogero Models}, 
hep-th/9812034.

\bibitem{hp}
P.S. Howe and G. Papadopoulos, Nucl. Phys. {\bf B381} 
(1992) 360, hep-th/203070.

\bibitem{mmt}
R. Minasian, G. Moore and D. Tsimpis, {\it Calabi-Yau Black Holes
and (0,4) Sigma Models}, hep-th/9904217.

\end{thebibliography}
\end{document}